\UseRawInputEncoding
\documentclass[twocolumn]{revtex4-1}

\makeatletter

\newcommand{\Rmnum}[1]{\expandafter\@slowromancap\romannumeral #1@}
\makeatother

\usepackage{amsthm}
\usepackage{mathrsfs}
\usepackage{amssymb}
\usepackage{epsfig}
\usepackage{amsmath}
\usepackage{graphicx}
\usepackage{epsfig}
\usepackage{booktabs}
\usepackage{multirow}
\usepackage[colorlinks,urlcolor=blue]{hyperref}
\usepackage{makecell}
\usepackage{bm}
\usepackage[utf8]{inputenc}
\usepackage{indentfirst}
\usepackage{color}

\begin{document}
\title{Ferroelectricity-Driven Metallicity and Magnetic Skyrmions in van der Waals Cr$_{2}$Ge$_{2}$Te$_{6}$/Hf$_{2}$Ge$_{2}$Te$_{6}$ Multiferroic Heterostructure}



\author{Zheng Chen}
\author{Hongliang Hu}

\affiliation{Zhejiang Key Laboratory of Quantum State Control and Optical Field Manipulation, Department of Physics, Zhejiang Sci-Tech University, Hangzhou 310018, China}

\author{Wenjun Zhang}
\affiliation{College of Physics and Information Engineering, Shanxi Normal University, Taiyuan 030031, China}

\author{Xiaoping Wu}
\affiliation{Zhejiang Key Laboratory of Quantum State Control and Optical Field Manipulation, Department of Physics, Zhejiang Sci-Tech University, Hangzhou 310018, China}

\author{Ping Li}
\affiliation{College of Science, Xinjiang Institute of Technology, Xinjiang 735400, China}


\author{Changsheng Song}
\email{cssong@zstu.edu.cn}
\affiliation{Zhejiang Key Laboratory of Quantum State Control and Optical Field Manipulation, Department of Physics, Zhejiang Sci-Tech University, Hangzhou 310018, China}
\affiliation{Longgang Institute of Zhejiang Sci-Tech University, Wenzhou, China}

\begin{abstract}

Two-dimensional (2D) multiferroic heterostructures present a promising platform for advanced spin devices by leveraging the coexisting ferromagnetic (FM) and ferroelectric (FE) orders. Through first-principles calculations and atomistic simulations, we reveal non-volatile control of metallicity and topological spin textures in the Cr$_{2}$Ge$_{2}$Te$_{6}$/Hf$_{2}$Ge$_{2}$Te$_{6}$ (CGT/HGT) heterostructure. Notably, manipulating ferroelectric polarization in HGT significantly modulates the magnetic anisotropy energy (MAE) and Dzyaloshinskii-Moriya interaction (DMI) of CGT/HGT, reversing the easy magnetization axis from in-plane to out-of-plane. By analyzing the atomic-resolved SOC energy ($\Delta E_{soc}$), it is found that the cause of the change comes from the Fert-Levy mechanism. Additionally, this polarization control enables the creation and annihilation of bimerons and skyrmions,  with interlayer sliding further altering magnetic ordering. Our findings offer valuable insights into magnetoelectric coupling and spin texture manipulation in 2D magnets, highlighting their potential for next-generation spintronic and memory devices.

\end{abstract}

\maketitle

\section{Introduction}
\label{sec:I}
Two-dimensional (2D) multiferroic materials, characterized by the coexistence of multiple ferroic orders such as ferromagnetism (FM) and ferroelectricity (FE) \cite{gao2021two,spaldin2010multiferroics,tang2019two}. have emerged as a promising platform for next-generation multifunctional electronic devices due to their inherent magnetoelectric coupling. This coupling enables the mutual modulation of magnetic and electric order parameters, offering significant potential for applications in nonvolatile memory and logic devices \cite{wang2003epitaxial,choi2009switchable,kimura2003magnetic}. Furthermore, the symmetry breaking inherent in multiferroic materials, coupled with  spin-orbit coupling (SOC), leads to the formation of a Dzyaloshinskii-Moriya interaction (DMI) \cite{PhysRevB.102.220409,PhysRevB.110.L060406,PhysRevLett.103.127201}. This interaction plays a crucial role in generating and stabilizing topologically protected spin textures such as magnetic skyrmions, which exhibit remarkable resilience against defects and thermal flutuations \cite{bogdanov1989thermodynamically,bogdanov1999stability,sampaio2013nucleation,2024arXiv240815974L}, as well as bimerons that are similarly endowed with protection in easy-plane magnets characterized by broken symmetry \cite{sun2020controlling,xu2020electric}. These spin textures hold great promise for high-density and low-power spintronic devices.

Despite the potential of multiferroic materials, single-phase multiferroics \cite{cheong2007multiferroics,wang2009multiferroicity,tokura2010multiferroics} often exhibit weak magnetoelectric coupling and predominantly exist in antiferromagnetic (AFM) ground states, which limits their ability to sustain topological spin textures. This challenge has redirected research focus towards 2D van der Waals (vdW) multiferroic heterostructures, where ferromagnetic (FM) and ferroelectric (FE) layers are combined to achieve indirect magnetoelectric coupling. In these heterostructures, nonvolatile control of electronic, magnetic properties, as well as topological spin textures in the FM layer can be achieved through ferroelectric switching in the FE layer.

Recent advancements have highlighted the potential for controlling magnetoelectric coupling in FM/FE heterostructures, offering new avenues for tunable spintronic devices. For instance, in the Cr$_{2}$Ge$_{2}$Te$_{6}$/In$_{2}$Se$_{3}$  \cite{gong2019multiferroicity} heterostructure, the reversal of polarization in the In$_{2}$Se$_{3}$ layer effectively switches the easy magnetization direction of Cr$_{2}$Ge$_{2}$Te$_{6}$, allowing for the electrical control of skyrmions  \cite{li2021writing}. Similarly, the Ge/Si-CrXTe$_{3}$(X=Si, Ge, and Sn) heterostructures exhibit strong interfacial coupling that induces significant inversion symmetry breaking, leading to enhanced DMI and the stabilization of skyrmions \cite{zhang2023generation}. In VBi$_{2}$Te$_{4}$/ In$_{2}$Se$_{3}$ heterostructures \cite{wang2023exploitable}, manipulation FE polarization triggers a phase transition from semiconductor to semimetal, demonstrating the versatility of these systems. Furthermore, bilayer CrI$_{3}$/In$_{2}$Se$_{3}$ \cite{yang2021realization} and CrI$_{3}$/Pt$_{2}$Sn$_{2}$Te$_{6}$ \cite{li2024nonvolatile} systems show that different polarization states can lead to distinct interlayer magnetic couplings. Additionally, the CrISe/In$_{2}$Se$_{3}$ \cite{shen2023manipulation} heterostructure reveals that vertical strain can effectively generate and annihilate skyrmions and bimerons without the need for an external magnetic field, showcasing the ability to control topological spin textures through ferroelectricity polarization reversion. However, challenges remain in fully understanding and controlling the magnetoelectric coupling in 2D multiferroic heterostructures, particularly in overcoming the vdW gap between layers, which can hinder effective interlayer coupling. The recently proposed 2D ferroelectric Hf$_{2}$Ge$_{2}$Te$_{6}$ \cite{jin2022intrinsically}, with its low polarization reversal barrier and structural compatibility with Cr$_{2}$Ge$_{2}$Te$_{6}$, offers a promising candidate for multiferroic heterostructures. Nonetheless, the nature of interlayer magnetoelectric coupling and its impact on electronic and magnetic properties, as well as the stability of topological spin textures, require further investigation.

In this study, we employ first-principles calculations and atomistic simulations to explore the non-volatile control of electronic, magnetic and the topological spin textures in a 2D FM/FE multiferroic Cr$_{2}$Ge$_{2}$Te$_{6}$/Hf$_{2}$Ge$_{2}$Te$_{6}$ (CGT/HGT) heterostructure. We demonstrate that polarization reversal in the HGT layer induces a transition from semiconductor to metallic behavior, and switches the easy magnetization axis from in-plane to out-of-plane. Additionally, the magnetic anisotropy energy (MAE) and DMI are shown to be tunable through ferroelectric polarization, primarily governed by the Te-$p$ orbitals of CGT. The ability to create and annihilate bimerons and skyrmions via polarization manipulation, coupled with the potential to reverse magnetic ordering through interlayer sliding, highlights the CGT/HGT heterostructure as a highly promising system for future spintronic and information storage applications.

\section{MODEL AND METHOD}
\label{II}
We perform first-principles calculations based on the density functional theory (DFT) implemented in the Vienna Ab initio simulation Package (VASP) \cite{kresse1996efficient}. The plane-wave energy cutoff is set to 500 eV for all calculations. The Perdew-Burke-Ernzerhof (PBE) \cite{PhysRevLett.77.3865} approximation is used to describe the exchange and correlation functional. To better describe the Cr and Hf 3$d$ electrons, the DFT+$U$ method is used to treat electron correlations, in which the effective Hubbard $U$ is set to 2.0 and 4.0 eV, respectively. A 20 Å vacuum layer is used to avoid interaction between adjacent layers. The systems are fully relaxed until the energy and residual force on each atom is less than $10^{-6}$ and 0.001 eV/Å, respectively. The Heyd-Scuseria-Ernzerhof (06)(HSE06) functional \cite{PAWAR2021113445} is employed to obtain the accurate band structures. The vdW correction with the Grimme (DFT-D2) method \cite{https://doi.org/10.1002/jcc.20495} is included in the structural optimization. A Monkhorst-Pack special $k$-point mesh \cite{PhysRevB.13.5188} of 6×6×1 for the Brillouin zone integration was found to be sufficient to obtain the convergence. The kinetic pathways of transitions between different polarization and stacked states were calculated using the climbing image nudge elastic band (CI-NEB) method \cite{PhysRevB.97.144104}. 
The magnetic properties of CGT/HGT heterostructure are investigated based on the following Hamiltonian:
\begin{equation}\label{eq1}
\begin{split}
H = &\sum\limits_{\left \langle i,j \right \rangle}J_{1}(\vec{S_{i}}\cdot\vec{S_{j}})
+ \sum\limits_{\left \langle k,l \right \rangle}J_{2}(\vec{S_{k}}\cdot\vec{S_{l}}) 
+ \sum\limits_{\left \langle m,n \right \rangle}J_{3}(\vec{S_{m}}\cdot\vec{S_{n}})\\ 
&+ \sum\limits_{\left \langle i,j \right \rangle}\vec{d_{ij}}\cdot(\vec{S_{i}}\times\vec{S_{j}})
+ K\sum_{i}(S_{i}^{z})^2 
+ \mu_{Cr} B\sum_{i}S_{i}^{z}. 
\end{split}
\end{equation}

The Heisenberg exchange coefficients $J_{1}$, $J_{2}$ and $J_{3}$ present the interaction between the nearest-neighbor (NN), second NN, and third NN of the magnetic Cr atom, respectively \cite{PhysRevB.106.094403}. The $\vec{d_{ij}}$ denotes the DMI vectors between spin $\vec{S_{i}}$ and $\vec{S_{j}}$. $K$ represents the single-ion anisotropy as well as $S_{i}^{z}$ represent the $z$ component of $\vec{S_{i}}$. The magnetic moment of the Cr atom is denoted by $\mu_{Cr}$ and $B$ represents the external magnetic field. Here, a negative value for $J$ indicates ferromagnetic (FM) coupling while while a positive value signifies antiferromagnetic (AFM) coupling. A positive $K > 0$ implies an out-of-plane easy axis whereas a negative value suggests an in-plane easy axis.  Additionally, the modulus length of $d_{ij} > 0$ ($d_{ij} < 0$), which corresponds to the energy difference of clockwise (CW) and anti-clockwise (ACW) spin configurations, reflects its the chirality of the DMI. The atomistic simulations are performed using Landau-Lifshitz-Gilbert (LLG) equation \cite{lLandau1935ONTT} by adopting the the calculated magnetic parameters of Eq. (1) based on the first principles calculations, and a large supercell containing 115200 sites with periodic boundary conditions is implemented in the Spirit package \cite{PhysRevB.99.224414} to explore the magnetic textures in CGT/HGT heterostructure.
\section{Results and discussion}
\label{III}
\begin{figure*}[!t]
	\begin{center}
		\includegraphics[width=14.0cm]{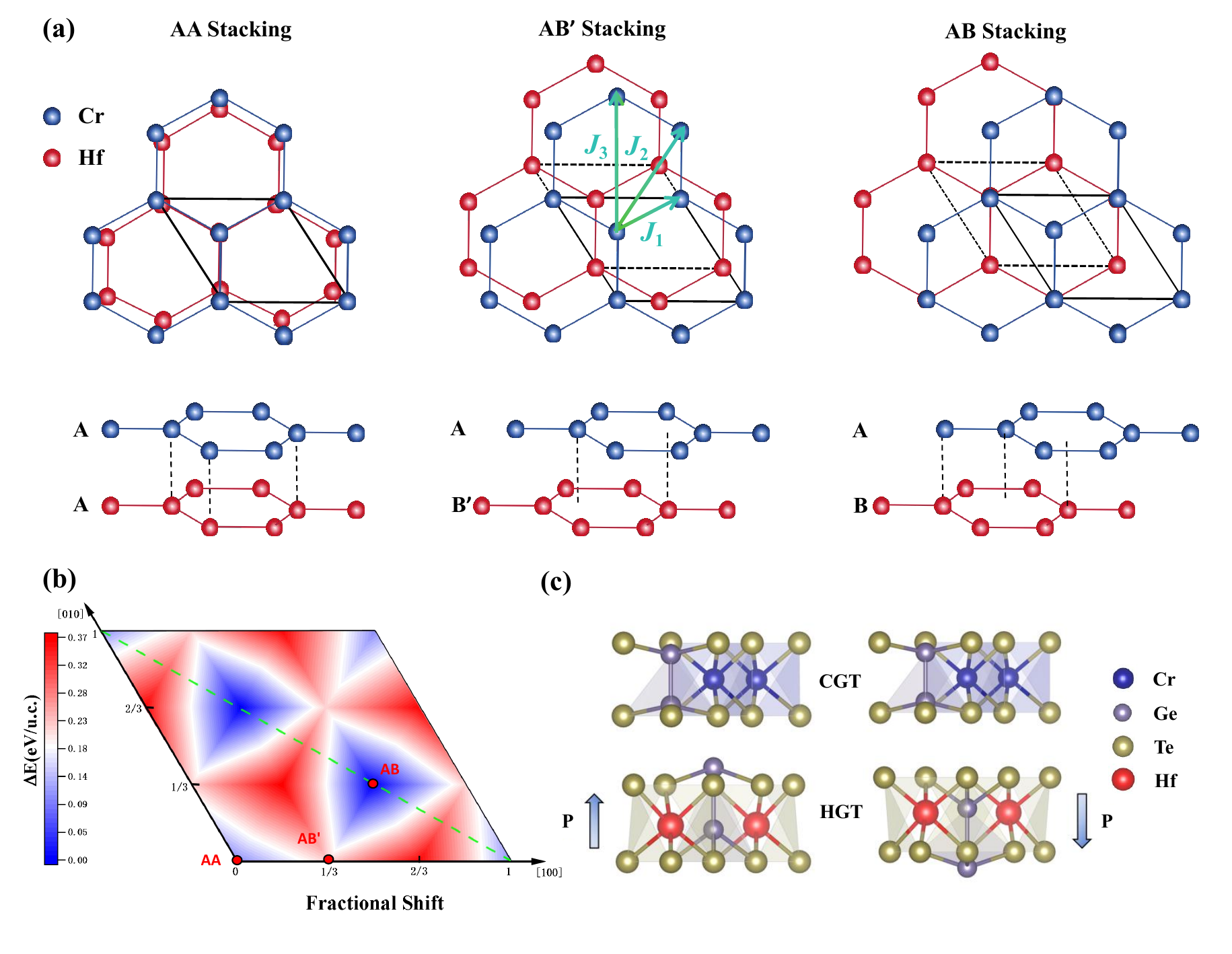}\
		\caption{\label{fig1}
(a) Schematic diagrams of AA, AB$'$ and AB stacking configurations in CGT/HGT heterostructure, where only the magnetic Cr and Hf atoms being displayed. The first, second, and third NN exchange-coupling parameters $J_{1}$, $J_{2}$, and $J_{3}$ are indicated respectively. (b) The energy landscape associated with sliding, where the green dashed line signifies sliding along the [$1 \overline{1} 0$] direction. (c) Crystal structures of CGT/HGT under AB stacking with ferroelectric polarization up (P$\uparrow$) and dw(P$\downarrow$) states.}
\end{center}
\end{figure*}

The heterostructure consists of a monolayer Cr$_{2}$Ge$_{2}$Te$_{6}$ (CGT) and a monolayer Hf$_{2}$Ge$_{2}$Te$_{6}$ (HGT), with lattice mismatch rate of 3.3 $\%$, where the individual optimized lattice constants of CGT and HGT are 6.91 and 7.14 Å, respectively. To simplify the stacking types of CGT/HGT heterostructures, only Cr and Hf atoms are retained and arrange in a honeycomb lattice, namely AA, AB$'$, and AB stacking in Fig. \ref{fig1}(a). To explore the energetically favored stacking orders, we calculate the stacking energy difference between AA-stacking and other stacking configurations. Starting with the AA-stacking, with the CGT layer kept fixed while shifting the HGT layer along the lattice vector [100] ([010]) by one-third of the lattice parameter to obtain nine stacking orders in Fig. \ref{fig1}(b). The corresponding energies are characterized by E$_{shift}$. The sliding energy $\Delta E_{sliding}$ can be defined as $\Delta E_{sliding}$= $E_{shift}-E_{AB}$. It is evident from our calculations that the AB stacking has the lowest energy, establishing it as the most stable configuration for our subsequent discussion. Meanwhile, AA stacking represents a metastable stacking structure, only slightly higher than AB stacking by 0.1 eV. Given the switchable polarization of HGT, illustrated in Fig. \ref{fig1}(c), two types of heterostructure naturally emerge: upward (up) and downward (dw) polarization direction respectively. In addition, for the AB-up configuration, the lattice constant is measured at 7.01 Å with an interlayer spacing is 2.44 Å. Whereas for AB-dw configuration, although there is no change in lattice constant, the interlayer spacing increases to 3.15 Å, significantly impacting both interlayer proximity effect and charge transfer mechanism that modulate the electronic and magnetic properties.

\begin{figure*}[!t]
	\begin{center}
		\centering
		\includegraphics[width=14.0cm]{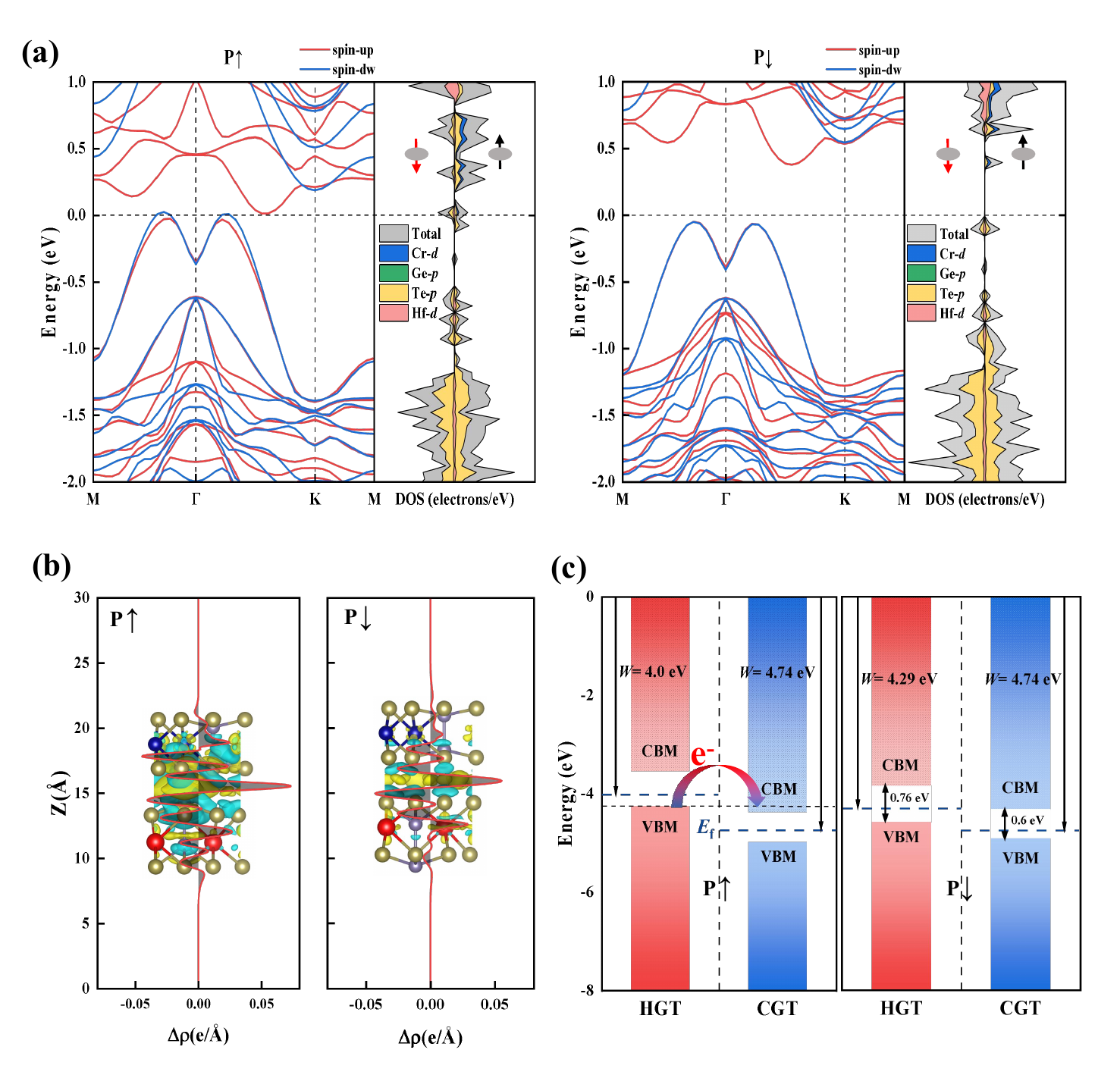}\
		\caption{\label{fig2}
(a) Band structures and projected density of states (PDOS) calculated by HSE06 hybrid functional, (b) plane-averaged differential charge density $\Delta \rho$(Z) and  charge density difference, (c) band alignments of pristine monolayer CGT and HGT in AB stacking CGT/HGT with ferroelectric polarization up and dw states.
}
\end{center}
\end{figure*}

The intrinsic HGT and CGT monolayer exhibits the semiconductor behaviors \cite{zhang2015robust,jin2022intrinsically}. The combination of the two compounds may lead to novel and significant discoveries, especially regarding the impact of polarization inversion in HGT on the electronic structure and magnetics characteristics in CGT/HGT multiferroic heterostructure. As shown in Fig. \ref{fig2}(a), the band structure of AB stacking CGT/HGT calculated by HSE06 hybrid functional exhibits a metallic behavior when ferroelectric polarization direction of HGT is oriented upwards (AB-up). Unexpectedly, when the polarization direction is flipped to downwards (AB-dw), the previous metallic behavior shifts towards semiconductor properties. Here,we also employ the PBE method to calculate the band for comparison and find the electronic structure near the Fermi level remains unchanged in Fig.S1 of Supplemental Material. Thus, we can conclude that the polarization reversal induces a transition from metallicity to semiconductivity in our CGT/HGT heterostructure. Given the more pronounced impact of ferroelectric polarization inversion on the electronic structure of the heterostructure compared to variations in stacking induced by interlayer sliding, the subsequent discussion primarily focuses on the influence of polarization inversion on magnetism under AB stacking.

To gain a more comprehensive understanding of the impact of polarization inversion in HGT on the electronic band structure of CGT/HGT heterostructure. As shown in Fig. S2 of Supplementary Material, we calculated the plane-averaged electrostatic potential of individual HGT and CGT monolayers. The structural asymmetric HGT exhibits a disparity in electrostatic potential along the out-of-plane direction, thereby resulting in a work function difference ($\Delta \varphi$) of 0.29 eV. In contrast, the spatial symmetry CGT displays an identical electrostatic potential on both sides without work function difference. The significant electrostatic potential difference between HGT and CGT induces a built-in electric field to make the redistribution of interface charge. As illustrated in Fig. \ref{fig2}(b), plane-averaged differential charge density $\Delta \rho$(Z) and  charge density difference clearly indicate much more charge exchange at the CGT/HGT interlayer with the AB-up state compared to that with the AB-dw stacking.

Based on the calculated band structures and plane-averaged electrostatic potential, the band alignment diagrams of AB-up and AB-dw in CGT/HGT are depicted in Fig. \ref{fig2} (c). The band structure of AB-dw stacking exhibits an indirect band gap, with the conduction band minimum (CBM) mainly contributed by the Cr-$d$ and Te-$p$ orbits of CGT, and the valence band maximum (VBM) coming from Te-$p$ orbit of HGT. It is observed that the energy level of VBM in isolated CGT is higher than that of HGT, while the energy level of CBM in CGT is higher than that of HGT, indicating a type-II band alignment for the CGT/HGT heterostructure, this type-II alignment can effectively separate electrons and holes, hindering charge transfer between CGT and HGT, thereby leading to a semiconductor properties. Conversely, when the polarization state of HGT switches to AB-up (P$\uparrow$ ), a more greater charge transfer occurs due to the VBM of HGT is energetically higher than the CBM of CGT, suggesting that the electrons in the valence band of HGT spontaneously transfer to the conduction band of CGT, resulting in metallic behavior.
 

\begin{figure*}[!t]
	\begin{center}
		\includegraphics[width=18.0cm]{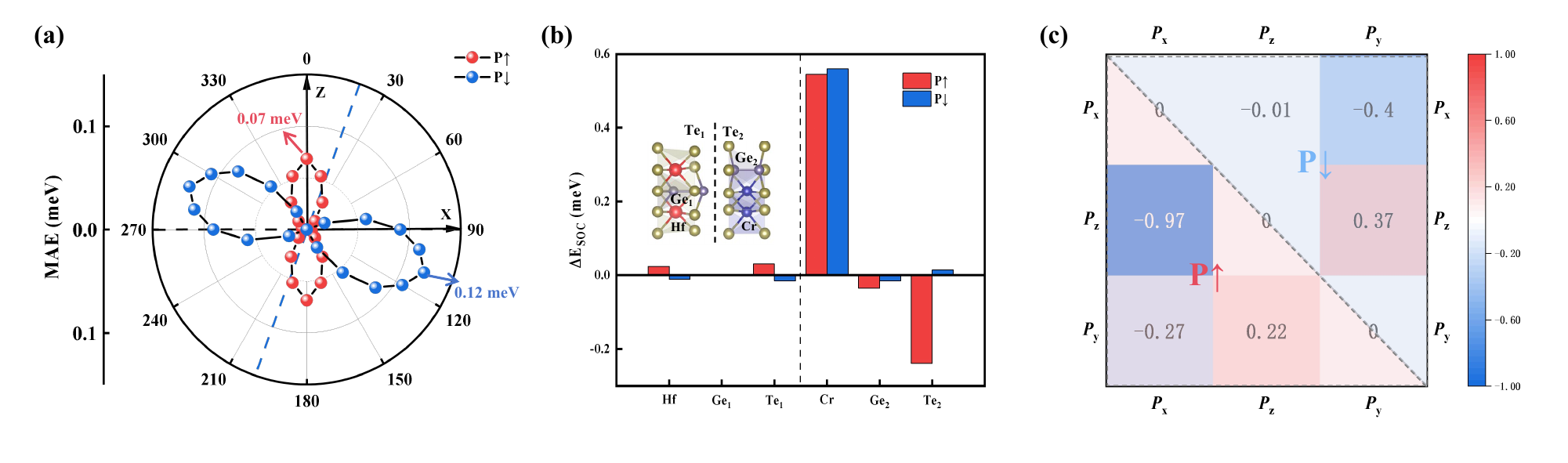}\
		\caption{\label{fig3}
(a) Angular dependence of MAE in the polar coordinate system and (b) element-resolved MAE of for CGT/HGT heterostructure with AB-up (red) and AB-dw (blue) states. The inset shows the side view of CGT/HGT heterostructure along with the corresponding atomic number distribution. (c) the Orbital-resolved MAEs of Te$_2$-$p$ orbitals with P$\uparrow$ and P$\downarrow$ states.
		}
	\end{center}
\end{figure*}

After the analysis above, we know that polarization inversion has a great influence on band structures and interfacial charge transfer. The following discussion will primarily focus on the impact of polarization inversion in HGT on the MAE within our HGT/CGT heterostructure. According to Mermin-Wagner theorem \cite{mermin1966absence}, a non-zero value of the MAE is the guarantee for a finite temperature and long-range magnetic order in 2D magnets. In addition, the low-temperature magnetic orientation with respect to the lattice structure and the thermal stability of spintronic devices is closely related to MAE, which the MAE is defined as MAE = $E_{x} - E_{z}$, where $E_{x}$ ($E_{z}$) is the energy of the unit cell when the spin points to the $x$ ($z$) direction. We know that previous theories predicted the MAE of 0.06 meV for monolayer CGT \cite{xue2023thickness}. Interestingly, the MAE of the CGT is significantly affected by the FE layer of HGT/CGT heterostructure, and the magnetic easy axis also undergoes rotation and deviation compared to that of the isolated CGT layer. The angular dependence of MAE in the polar coordinate system is shown in the Fig. \ref{fig3}(a), with the dotted lines indicating the direction of the magnetic easy axis. The calculated MAE of AB-up stacking of HGT/CGT is 0.07 meV, demonstrating a tendency towards alignment with X-axis and isotropy within the XY in-plane (See Fig. S5 of Supplementary Material). However, for AB-dw stacking, the MAE of CGT reaches up as high as 0.12 meV, and resulting in out-of-plane anisotropy with a $20^{\circ}$ rotation relative to the Z axis. This can be attributed to dominance of Cr-3$d$ electrons$’$ orbitals with Z-component in an octahedral field, clearly indicating that distortion induced rearrangement of the 3$d$ electrons is responsible for such slant easy axis \cite{chen2021two}. Therefore, by controlling the ferroelectric polarization direction of the HGT layer, a switching of magnetic easy axis can be achieved. This capability is beneficial for applications involving the storage and manipulation of magnetic information.

To obtain the origin of inversion of MAE induced by the polarization direction of FE layer in our CGT/HGT heterostructure, as shown in Fig. \ref{fig3}(b), we have calculated the element-resolved MAEs of the with AB-up and AB-dw states. The MAE of the isolated CGT mainly originates from the Cr and Te atoms \cite{liu2022magnetic}. After forming the heterostructure, the element-resolved MAEs show that the switching behavior of MAE is mainly determined by the difference of Te$_2$ rather than the Cr atoms with the consideration of the up or down polarization direction. Furthermore, we calculate the orbit-resolved MAE of the Te$_2$-$p$ orbitals as illustrated in Fig. \ref{fig3}(c). For AB-up stacking, the hybridization between the $p_x$ and $p_z$ orbitals mainly contribute to the in-plane magnetic anisotropy (IMA), while the hybridized $p_y$ and $p_z$ orbitals contributes significantly to the perpendicular magnetic anisotropy (PMA). The competition between the hybridized $p_x$-$p_z$ and $p_y$-$p_z$ leads to the small IMA (MAE =$-0.07$ meV ) for CGT/HGT heterostructure with AB-up state. However, when the polarization is reversed to AB-dw as demonstrated in Fig. \ref{fig3}(d), the hybridization coulping of $p_x$ and $p_z$ orbitals enhanced toward the trend of PMA by approximately 0.96 meV, while the hybridized $p_y$-$p_z$ with a small growth of 0.15 meV for IMA. The stably increasing trends of PMA is much larger than that of IMA, resulting in the positive MAE = 0.12 meV for CGT/HGT heterostructure with an off-plane magnetization axis. 

\begin{figure*}[!t]
	\begin{center}
	\includegraphics[width=18.0cm]{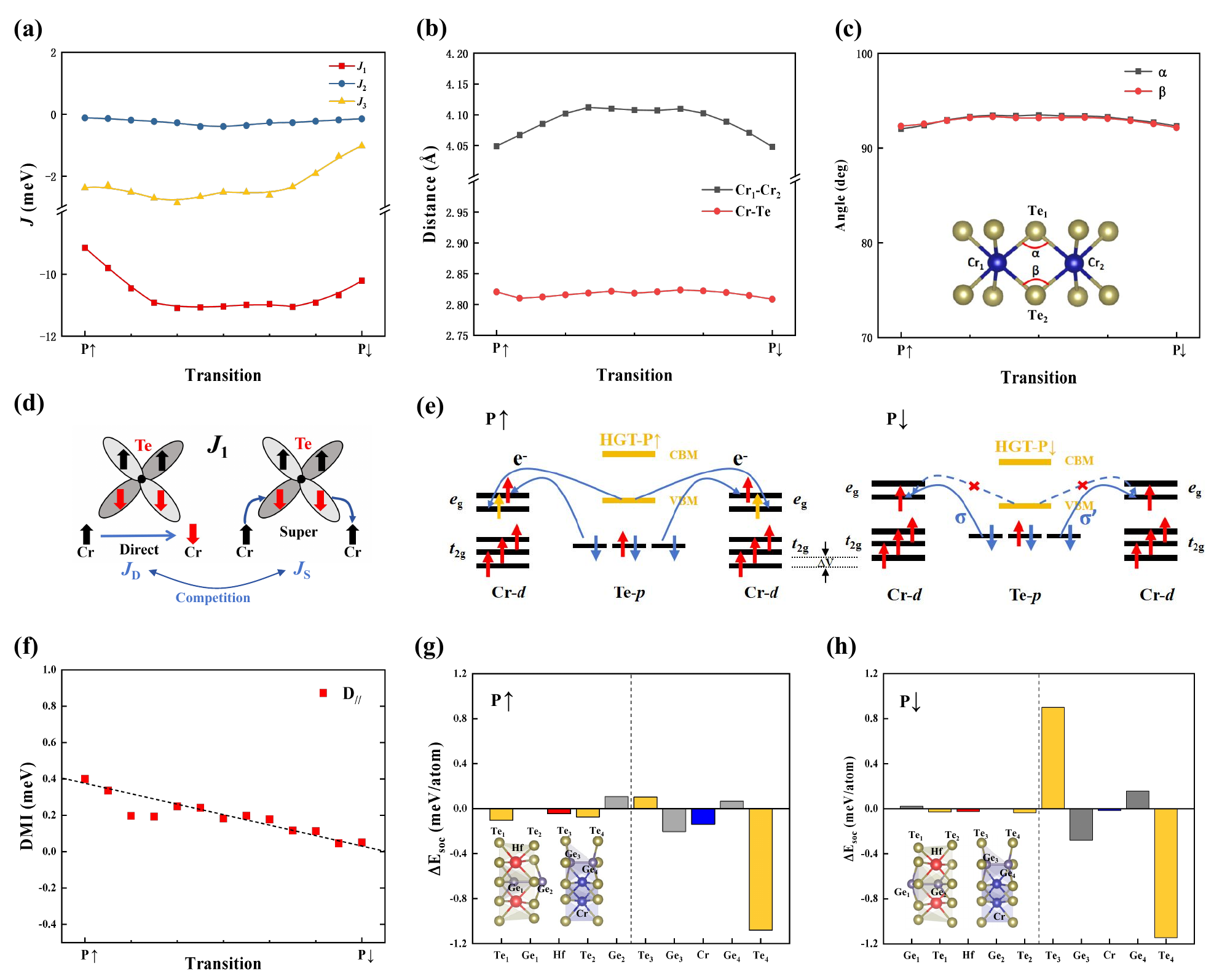}\
	\caption{\label{fig4}
The change curve of heisenberg exchange coupling ($J_{1}$, $J_{2}$, $J_{3}$) in CGT/HGT heterostructure during the evolution from P$\uparrow$ to P$\downarrow$. (b) The bond lengths of Cr$_{1}$-Cr$_{2}$ and Cr-Te in CGT, as well as (c) the bond angles $\alpha$ and $\beta$, are analyzed as a function of the ferroelectric transition. (d) Schematic of Heisenberg exchange interactions $J_{1}$, caused by the competition between direct exchange and hyperexchange interactions between Cr-Cr orbitals. (e) Schematic diagram of intralayer (near-$90^{\circ}$) Cr-Te-Cr super-exchange interaction under P$\uparrow$ and P$\downarrow$ stacking. (f) The intensity of DMI as functions of ferroelectric polarization reversing in AB stacking. Atomic-resolved SOC energy ($\Delta E_{soc}$) associated with DMI  under (g) AB-up and (h) AB-dw stacking. The insets show the atomic number distribution of CGT/HGT heterostructure.
	}
	\end{center}
	\end{figure*}

As mentioned above, the reversal of the polarization direction can manipulate the magnetization axis of the CGT. However, it is currently unclear how the other microscopic magnetic parameters, such as Heisenberg exchange coupling $J$ and DMI in HGT/CGT heterostructure, change during the process of reversing the polarization direction. Herein, we utilize the nudged elastic band (NEB) method to investigate this issue. Firstly, we examine the variation of heisenberg exchange coupling ($J_{1}$, $J_{2}$, $J_{3}$) in CGT/HGT heterostructure from AB-up to AB-dw states. As depicted in Fig. \ref{fig4}(a), $J_{1}$ and $J_{3}$ undergo an initially increasing and then decreasing, whereas $J_{2}$ plays a less important role due to minimal Te-Te hybridizations, being more than an order of magnitude smaller than $J_{1}$ and $J_{3}$. Therefore, from the numerical point of view, the ferromagnetism of HGT/CGT system is mainly determined by the size of $J_{1}$. To further explore the law of the variation $J_{1}$ from the view of microphysical mechanisms, as depicted in Fig. \ref{fig4}(b) and (c), we analyze the variation of the Cr$_{1}$-Cr$_{2}$ and Cr-Te bond lengths, as well as the Cr$_{1}$-Te$_{1}$-Cr$_{2}$ ($\alpha$) and Cr$_{1}$-Te$_{2}$-Cr$_{2}$ ($\beta$) bond angles. The results show that the distance ($d_{Cr_{1}-Cr_{2}}$) between Cr$_{1}$ and Cr$_{2}$ increases and then decreases when transitioning from AB-up to AB-dw stacking, implying that the direct-exchange interaction ($J_{D}$) of antiferromagnetic coupling decreases and then increases. However, the super-exchange interaction ($J_{S}$) of ferromagnetic coupling is basically unchanged because the bond length (Cr-Te) and the bond angle of CGT changes little. Therefore, as shown in Fig. \ref{fig4}(d), according to the Goodenough-Kanamori-Anderson (GKA) rules \cite{goodenough1955theory,kanamori1959superexchange,anderson1959new}, because the $J_{S}$ dominates the competition with the $J_{D}$, which leads to the $J_{1}$ of the CGT being ferromagnetically coupled ($J_{1}<0$) . Therefore, the change in the magnetic coupling strength when transitioning from AB-up to AB-dw is due to the ferroelectric inversion makes the heterojunction have different degrees of interlayer coupling, changing the distance between Cr$_1$ and Cr$_2$, resulting in a change in the direct exchange interaction of the antiferromagnetic coupling, which in turn makes $J_{1}$ change.

In theory, AB-up and AB-dw states with similar bond lengths and bond angles should exhibit comparable magnetic coupling strengths. However, as depicted in Fig. \ref{fig4}(a), the magnetic coupling strength of AB-dw is approximately 1 meV higher than that of AB-up. This is illustrated in in Fig. \ref{fig4}(e) with the orbital coupling between Cr-$d$ (short black line) and HGT (short yellow line), along with a virtual orbital hopping path (long blue line) between two Cr-$t_{2g}$ orbitals already occupied by three spin-up electrons. In AB-dw heterostructure characterized by a CGT octahedral arrangement, the $t_{2g}$ orbitals are strongly hindered from forming $\sigma$ ($\sigma$$'$) bonds with the anion Te \cite{watson2020direct} when the Cr-Te-Cr bond angle approaches $90^{\circ}$. As a result, there is no mechanism to link the spin orientations of the $t_{2g}$ shells between neighboring Cr sites. However, the $p$-$e_{g}$ orbital hybridization leads to electron occupying the eg shell, aligning parallel to the local spin direction of $t_{2g}$ at the Cr site due to Hund's rule coupling \cite{kutzelnigg1996hund}. This process creates a hole in the corresponding Te-$p$ orbital, facilitating a natural pathway for the superexchange mechanism between two Cr atoms. Conversely, in AB-up heterostructure, the ferroelectric HGT monolayer aligns with type-III band alignments with CGT. The valence band maximum (VBM) of HGT shifts up into the conduction band minimum (CBM) of Cr-$e_{g}$, causing the Cr-$e_{g}$ orbitals to undergo charge transfer from the VBM of HGT. This results in $p$-$e_{g}$ orbital hybridization that necessitates higher energy to facilitate orbital hopping between Cr-$t_{2g}$ orbitals. Consequently, this inadvertently weakens the strength of the superexchange ferromagnetic coupling mediated by Te. Moreover, the AB-up configuration induces a positive E$_{eff}$ that represents the effective electric field crossing the CGT/HGT heterostructure, which also reduces the potential energy ($\Delta$V) of Cr-$d$ states and diminishes the strength of $p$-$e_{g}$ orbital hybridization, thereby suppressing virtual hopping between two Cr-$t_{2g}$ orbitals. Therefore, ferroelectric proximity effect causes different degree of charge transfer between the intrelayer, changes the virtual orbital hopping path between the two Cr-$t_{2g}$ orbits, and thus changes the ferromagnetic coupling of the superexchange.

\begin{figure*}[!t]
	\begin{center}
		\includegraphics[width=18.0cm]{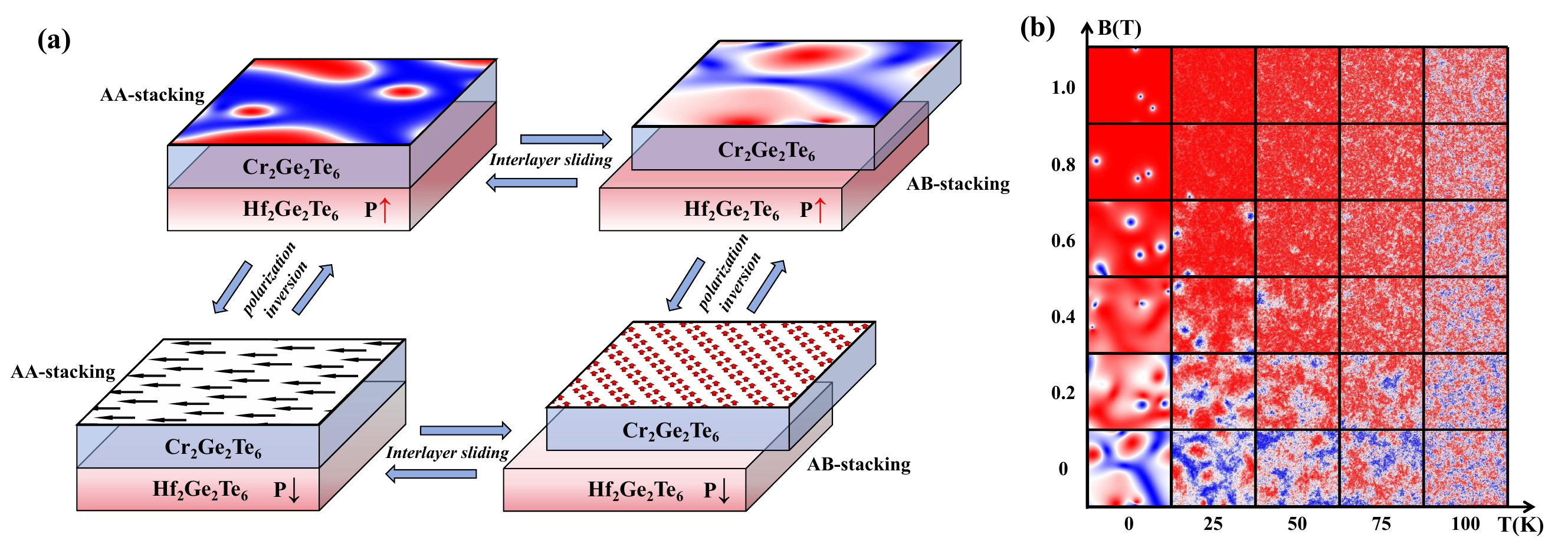}\
		\caption{\label{fig5}
(a) Schematic diagram illustrating the manipulation of topological spin textures and magnetic ordering through ferroelectric polarization and interlayer sliding. (b) The evolutions of spin textures of the HGT/CGT heterostructure with different temperature and external magnetic field at P$\uparrow$ stacking.
		}
	\end{center}
\end{figure*}

The DMI typically arises from the breaking of spatial inversion symmetry and SOC. It plays a crucial role in the formation of topological spin textures, such as skyrmions and bimerons. As illustrated in Fig. \ref{fig4}(f), the DMI in the CGT/HGT heterostructure varies across different polarization configurations, ranging from AB-up to AB-dw, with a nearly linear decrease from 0.4 meV to 0.07 meV. On the one hand, we think that the change primarily results from structure deformation when reversing ferroelectric polarization. As shown in Fig. S6, when the ferroelectric polarization is reversed, the CGT in the heterostructure can be found a remarkable structural deformation which shows the polarization dependence of verical distance (z-component) between distance of Cr-Te$_3$ ($d_{Cr-Te_{3}}$) and Cr-Te$_4$ ($d_{Cr-Te_{4}}$). We can find that the Cr atoms gradually move closer to the Te$_3$ layer from their original proximity to the Te$_4$ layer. Such a large geometrical change is due to the different polarization structure of HGT, which introduce sizable atomic force. Similar phenomena and mechanistic explanations have been reported previously \cite{RN3110}. On the other hand, originating from the strong SOC interactions of Te atoms, the Fert-Levy mechanism \cite{Fert1980RoleOA,fert1990magnetic} generates significant DMI through the Cr-Te-Cr exchange path. Under the combined influence of the ferroelectric polarization, as depicted in Fig. \ref{fig4}(g) and (h), we analyze the microscopic mechanism of DMI in the AB-up and AB-dw configurations by calculating the atomic-resolved SOC energy ($\Delta E_{soc}$). We find that a stronger hybridization effect occurs in the interlayer Te$_3$ atoms of the AB-up structure compared to AB-dw, resulting in a significant reduction of $\Delta E_{soc}$. It can also be seen from Fig. \ref{fig2}(b) that the transfer of electrons and orbital hybridization of Te$_3$ atoms at the interface of AB-up stacking is stronger compared to AB-dw stacking. Thus, ferroelectric inversion can significantly modulate the magnitude of the DMI, paving the way for the subsequent production and annihilation of topological spin textures.

Finally, utilizing the magnetic parameters obtained from the first principles calculations, we conduct atomistic simulations to investigate the revolution of the magnetic configurations of HGT/CGT heterostructure through interlayer sliding and FE switching manipulation. As shown in Fig. \ref{fig5}(a), starting with the AA-up stacking, we identified an intrinsic topological skyrmion state. Upon sliding the HGT layer to AB-up, the spin texture behaves as bimerons due to easy-plane anisotropy. Interestingly, the skyrmions in AA-up stacking annihilate to display an in-plane ferromagnetic ordering in AA-dw stack and the bimerons AB-up stack annihilate to display an out-plane ferromagnetic ordering in AB-dw stack when the HGT ferrielectric polarization switches to the opposite direction. This illustrates that the creation and annihilation of topological spin textures can be achieved via electrically driven magnetism in 2D multiferroic heterostructures \cite{PhysRevB.102.014440,PhysRevResearch.3.L012026,PhysRevB.107.184439}. Additionally, interlayer sliding proved effective in reversibly controlling different spin textures \cite{PhysRevB.109.024426,10.1063/5.0190515}. It is noteworthy that the magnetic field and temperature significantly influence chiral spin textures in 2D magnets \cite{wang2020independent,du2022spontaneous} and that the topological spin textures annihilate under a large magnetic field and high temperature. Thus, we further investigate the evolutions of spin textures of the AB-up HGT/CGT heterojunction under temperature and external magnetic fields. As shown in Fig. \ref{fig5} (b), applying an in-plane magnetic field at 0K gradually reduces the size of bimerons. When the magnetic field is increased to 0.6T, the spin texture transforms into skyrmions, indicating a topological phase transition. Furthermore, applying a 0.4T out-of-plane magnetic field at 50K still maintained distinct spin textures. As the temperature increased further, the magnet transitions from its original ferromagnetic state to a paramagnetic phase, and the spin texture becomes disordered. For more detailed spin textures in AB-up and dw stacking, refer to Supplementary Fig. S7. Our study presents a promising strategy for achieving nonvolatile control over various magnetic states, demonstrating potential for generating novel spin textures.

\section{Conclusions}
\label{IV}
In conclusion, by first-principles calculations and atomistic simulations, we have demonstrated that the vdW CGT/HGT multiferroic heterostructure serves as an ideal platform for nonvolatile control of both metallic properties and topological spin textures. Our results reveal that the reversal of ferroelectric polarization induces significant changes in the piezoelectric field and enhances interlayer charge transfer within the CGT/HGT heterostructure, leading to a transition from a semiconductor to a metal, driven by type-II and type-III band alignments. We further elucidate the underlying mechanisms of the reduction in magnetic coupling strength due to polarization reversion, attributing it to the interplay between direct exchange and super-exchange interactions. Notably,  electron transfer from Te-$p$ orbitals in HGT to Cr-$e_{g}$ orbitals in CGT along an orbital hopping path weakens the superexchange strength, complementing the direct exchange in reducing ferromagnetic coupling. Additionally, we demonstrate that the switchable MAE and the near-linearly decreasing DMI induced by ferroelectric polarization reversal are primarily governed by the Te-$p$ orbitals of CGT. Furthermore, the ability to create and annihilate bimerons and skyrmions through polarization manipulation, along with the capability to reverse magnetic ordering and alter the easy magnetization axis via interlayer sliding, underscores the significant potential of this heterostructure for future spintronic and memory device applications. This study not only enhances our understanding of multiferroic heterostructures but also provides valuable insights for  the design of spintronic with robust electromagnetic control.

\begin{acknowledgments}
We thank Dongzhe Li for valuable discussions and constructive comments. This work was supported by National Natural Science Foundation of China (Grant No.11804301), the Natural Science Foundation of Zhejiang Province (Grant No.LY21A040008), the Fundamental Research Funds of Zhejiang Sci-Tech University (Grant Nos.2021Q043-Y and LGYJY2021015), the Excellent Postgraduate Dissertation Cultivation Funds of Zhejiang Sci-Tech University (Grant LW-YP2024013).
\end{acknowledgments}

\bibliography{ref}
\end{document}